\documentclass[
reprint,
amsmath,amssymb,
aps,prl
]{revtex4-1}

\usepackage{graphicx}
\usepackage{dcolumn}
\usepackage{subfigure}
\usepackage{hyperref}
\usepackage{siunitx}
\usepackage[final]{changes}

\newcommand{\Rey}{\mathit{Re}}

\begin{document}

\title{{Dispersed fibers change the classical energy budget of turbulence
via nonlocal transfer}}

\author{Stefano Olivieri$^{1,2,3}$}
\author{Luca Brandt$^4$}
\author{Marco E. Rosti$^3$}
\email[Corresponding author:]{marco.rosti@oist.jp}
\author{Andrea Mazzino$^{1,2}$}
\affiliation{
 $^1$ Department of Civil, Chemical and Environmental Engineering (DICCA), University of Genova, Via Montallegro 1, 16145, Genova, Italy \\
 $^2$ INFN, Genova Section, Via Montallegro 1, 16145, Genova, Italy \\
 $^3$ Complex Fluids and Flows Unit, Okinawa Institute of Science and Technology Graduate University, 1919-1 Tancha, Onna-son, Okinawa 904-0495, Japan\\
 $^4$ Linn\'{e} FLOW Centre and SeRC, Department of Engineering Mechanics, KTH Royal Institute of Technology, Stockholm, Sweden
}

\date{\today}

\begin{abstract}
The back-reaction of dispersed \added{rigid fibers} to turbulence is analyzed by means of a state-of-the-art fully-coupled immersed boundary method. The following universal scenario is identified: turbulence at large scales looses a consistent part of its kinetic energy (via a Darcy friction term), which partially re-appears at small scales where a new range of energy-containing scales does emerge. Large-scale mixing is thus depleted in favor of a new mixing mechanism arising at the smallest scales. Anchored fibers cause the same back-reaction to turbulence as moving fibers of large inertia. Our results thus provide a link between two apparently separated realms: the one of porous media and the one of suspension dynamics.
\end{abstract}

\pacs{Valid PACS appear here}

\maketitle

The interaction between fluid flows and dispersed objects concerns a wide range of physical problems with both environmental and industrial application, such as transport processes in canopies or porous media, as well as suspension dynamics and complex fluids~\cite{finnigan2000review,nepf2012review,butler2018review,apte2020review}. 
In boundary-layer meteorology, e.g., the presence of plant canopies modifies the momentum and heat fluxes, consequently altering ecological mechanisms of primary importance such as carbon dioxide exchange~\cite{raupach1981,finnigan2000review,nepf2012review,lemone100years}, with important implications on climate changes. 
In this framework, Ref.~\cite{finnigan2000review} highlighted how the classical turbulence scenario well explained by Kolmogorov's theory~\cite{frisch1995turbulence} can be intrinsically modified by the presence of the canopy: the drag exerted by the latter causes the energy of the flow to be extracted at the large scales and partially reintroduced at smaller scales where fine flow structures are generated. This energy transfer mechanism, labeled as `spectral shortcut', is commonly invoked to explain why the Kolmogorov scaling for the energy spectrum $E(k) \sim k^{-5/3}$ does not hold in canopy turbulence~\cite{finnigan2000review,lemone100years}.
However, the full comprehension of this phenomenon on a more fundamental basis is still missing and object of active research~\cite{poggi2004,DiBernardino2017,ghisalberti_nepf_2009a,zampogna_bottaro_2016,zampogna2016,shnapp2019extended,monti2019PoF}.
Similarly, for the case of suspensions, the back-reaction to the flow due to the presence of dispersed particles is poorly understood and no fully predictive models exist~\cite{saffman1962,dodd_ferrante_2016,lucci_ferrante_elghobashi_2010,bec2017dusty,gualtieri2017turbulence,rosti_ge_jain_dodd_brandt_2019}.
	
\begin{figure}[b]
    \centering
    \includegraphics{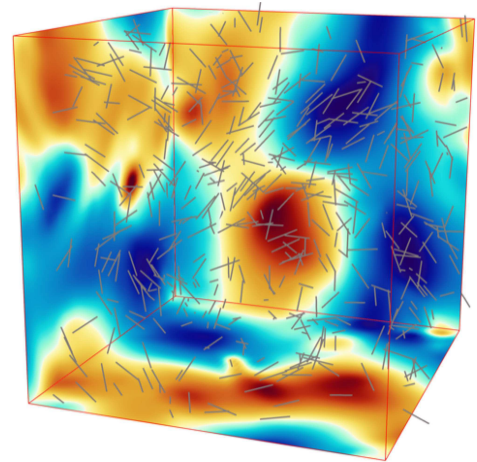}
    \caption{
    A snapshot of the three-dimensional flow under investigation and $N = 8^3$ fibers of \added{length $c/L = (4 \pi)^{-1}$} dispersed within. Cutplanes are coloured by the pressure field.}
    \label{fig:intro}
\end{figure}

In this Letter, we reveal the existence of a nonlocal energy transfer mechanism universally present when a turbulent flow interacts with a dispersed phase.  
To describe this multiscale fluid-structure interaction problem, we consider a turbulent flow with dispersed \added{rigid fibers}, as shown in Fig.~\ref{fig:intro}. 
We present results of a scale-by-scale spectral analysis showing that i) the large-scale dynamics can be effectively described in terms of a Darcy friction (when the flow is observed on scales much larger than the typical correlation length of the dispersed phase), and ii) the lengthscale at which energy is reintroduced in the flow is the characteristic distance between the dispersed elements, and not their size as usually assumed.

The problem is tackled by means of direct numerical simulations (DNS) complemented with an immersed boundary method (IBM)~\cite{huang_shin_sung_2007a,rosti2018flexible,rosti2019flowing}~\footnote{For more information on the numerical approach, the reader is referred to the Supplementary Materials} to comprehensively investigate both the small- and the large-scale dynamics. 
In particular, we consider a three-periodic fluid domain of size $L=2\pi$ where turbulence is sustained by forcing the incompressible Navier-Stokes equations at the largest scale (i.e., on the first wavenumber $k=1$), in the unstable regime of the Arnold-Beltrami-Childress (ABC) flow at $\Rey \equiv \nu^{-1} = 130$, being $\nu$ the kinematic viscosity of the fluid~\citep{dombre1986chaotic,podvigina1994}.
An ensemble of $N$ elongated \added{rigid fibers of length $c/L =(4\pi)^{-1}$} are immersed in the flow 
and their concentration can be quantified by the number density $n = N/ L^3$;
the latter is always such that we fall into the dilute regime, i.e. $n \, c^3 \ll 1$~\cite{butler2018review}. 
To better isolate the different effects, fibers can be either fixed or moving. 
In the fixed configuration, we consider different fiber lengths $c$ and concentrations $n$, isotropic and anisotropic orientation distributions and both evenly and randomly spaced fibers.
In the moving configuration, we vary the fiber inertia by changing the linear density difference between the fiber and the fluid, $\Delta \widetilde{\rho}$. 

\begin{figure}
    \centering
    \includegraphics{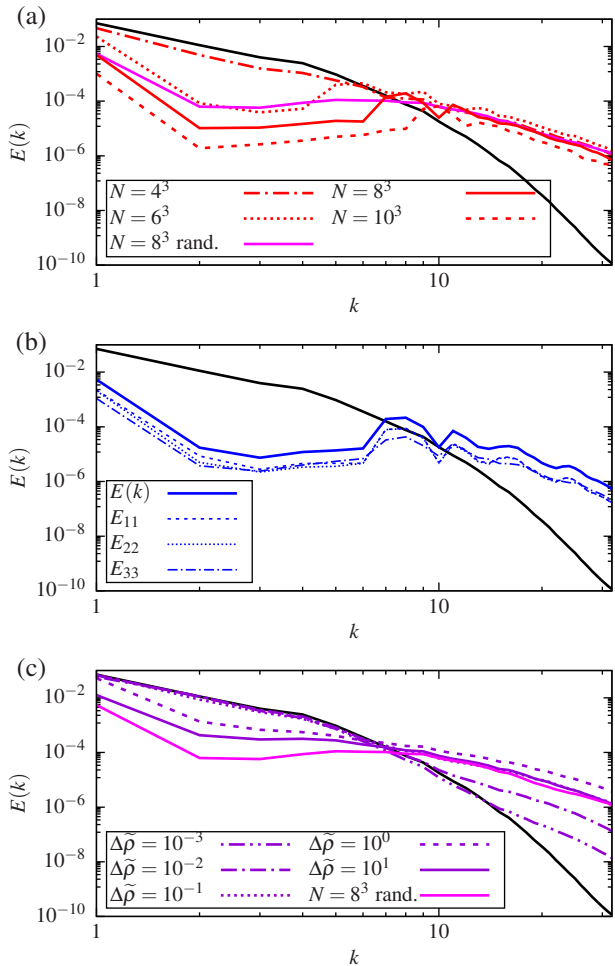}
    \caption{
    Energy spectra for (a) fixed fibers with isotropic orientation for different fiber concentrations, (b) fixed fibers evenly spaced with anisotropic orientation, (c) moving fibers with different inertia (along with the magenta curve for randomly-distributed fixed fibers reported again for comparison). The black curve represents the single-phase case, i.e. $N=0$.
   }
    \label{fig:spectra}
\end{figure}
	
Fig.~\ref{fig:spectra} shows the energy spectra $E(k)$ for several cases from our numerical study. First, results for the fixed-fiber configuration with isotropic orientation are collected in Fig.~\ref{fig:spectra}a. 
Starting from the single-phase solution (i.e., $N=0$) and increasing the fiber concentration, the amplitude of the energy-containing low-wavenumber components is found to decrease. 
The presence of fibers alters the energy distribution across the scales of motion: while the large-scale/low-wavenumber components are damped, the small-scale/high-wavenumber activity is enhanced. 
Although the first wavenumber remains the dominant mode, a secondary peak is clearly observed at a higher wavenumber $k_\mathrm{c} = 2\pi / \ell$. We find that the associated lengthscale $\ell$ corresponds to the spacing between fibers, i.e. $\ell = L / \sqrt[3]{N}$. 
Notice that the characteristic length of the single canopy element (in our case the fiber length $c$) does not appear directly, in contrast to what typically claimed for turbulence in plant canopies~\cite{lemone100years,finnigan2000review}.
Fig.~\ref{fig:spectra}a also includes the results for fibers randomly distributed in the domain: a similar behavior is evident, except for a smoother transition in the intermediate range of scales. This difference is due to the uneven spacing between fibers, resulting in a distribution of $k_\mathrm{c}$ instead of a uniquely defined value.
A similar outcome is obtained when the fiber orientation has an anisotropic distribution, as reported in Fig.~\ref{fig:spectra}b. Here, fibers lie on evenly spaced planes and have a two-dimensional orientation distribution. In this case, the weakening of the flow due to fibers is less intense for the velocity components parallel to the planes ($E_{11}$ and $E_{22}$) than for the normal component ($E_{33}$). Nevertheless, the structure of the energy spectra is the same, with a reduced energy level for large scales and a secondary peak emerging at the same characteristic lengthscale $\ell$ previously identified.
Finally, Fig.~\ref{fig:spectra}c shows the resulting spectra for the configuration where fibers are moving.
The feedback of fibers to the flow causes energy reduction at large scales associated to an enhancement occurring at small scales. While the large-scale suppression becomes irrelevant in the limit of vanishing inertia, this is not the case for the small-scale intensification. This indeed appears even for negligible fiber inertia when the fiber and fluid velocities are very close to each other. In this limit fibers only act to increase the effective fluid density, thus increasing the effective Reynolds number of the flow and, ultimately, the resulting small-scale turbulence. This mechanism is the same at work for fine dust in a flow field~\cite{saffman1962,bec2017dusty}.
The large-scale suppression increases as the fiber inertia becomes more and more important giving rise to a spectral gap.  For the largest fiber inertia we have analyzed, the energy distribution along the scales is very similar to that of the fixed randomly-distributed fibers having the same number density (see the magenta curve in Fig.~\ref{fig:spectra}c).
As we will see in the following of this Letter, large-scales are subject to a Darcy dissipation mechanism. However, for the latter dissipative effect to emerge, sufficiently large velocity differences between fibers and fluid must occur.
\added{Note that, the underlying mechanism here is substantially different from those at play in turbulent flows with polymer additives that are typically inertialess, elastic and of microscopic size~\cite{de_angelis2005,boffetta2005drag,berti2006,nguyen2016}.}
\begin{figure}
    \centering
    \includegraphics{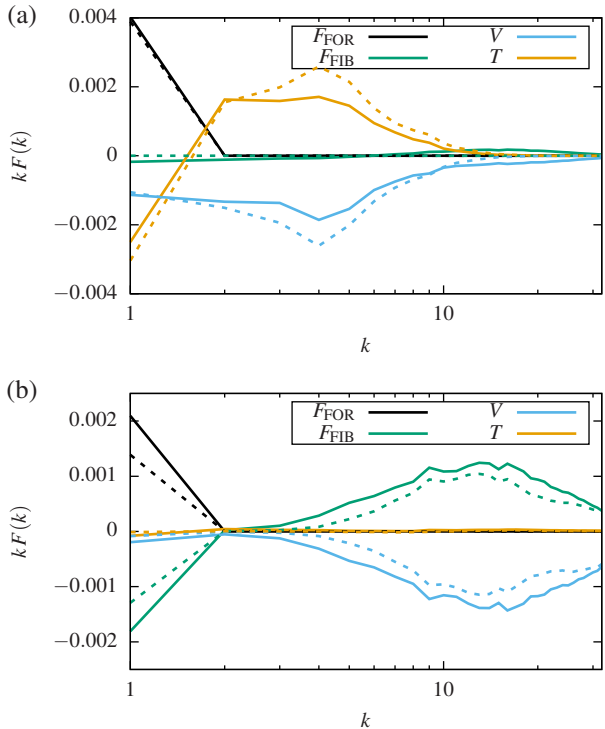}
    \caption{Spectral energy balance according to Eq.~\eqref{eq:loads} for 
    (a) $8^3$ moving fibers of length $c/L= (4 \pi)^{-1}$ with small inertia ($\Delta \widetilde{\rho} = 10^{-2}$, solid curves), compared to the single-phase case (i.e., $N=0$, dashed curves);
    (b) the same configuration but with large inertia ($\Delta \widetilde{\rho} = 10^{1}$, solid curves), compared to the case where fibers are fixed with random position and isotropic orientation (dashed curves). All the quantities are premultiplied by $k$ to improve the plot readibility. Black: external forcing $F_\mathrm{FOR}$; green: fluid-structure coupling $F_\mathrm{FIB}$; cyan: viscous dissipation $V$; orange: nonlinear term $T$.}
    \label{fig:loads_ABC}
\end{figure}

To get a deeper understanding of how energy is redistributed in the flow, we consider the governing equation for the energy spectrum
\begin{equation}
        \partial_t {E} =  T + V + F_\mathrm{FOR} + F_\mathrm{FIB},
    \label{eq:loads}
\end{equation}
where the quantities appearing on the right-hand-side correspond to the nonlinear energy transfer $T$, the viscous dissipation $V$, the external flow forcing $F_\mathrm{FOR}$ and the fluid-structure coupling $F_\mathrm{FIB}$~\footnote{For the definition of these quantities, see the Supplementary Materials.}.
For steady or statistically steady flows, the left-hand-side becomes zero, and the energy balance is governed only by the four terms appearing on the right hand side.
In Fig.~\ref{fig:loads_ABC}, each of these terms is plotted as a function of the wavenumber in the two limiting situations outlined previously. 
We first consider the case of fibers with relatively small inertia (Fig.~\ref{fig:loads_ABC}a). Here, the fluid-structure coupling $F_\mathrm{FIB}$ turns out to be negligible at the lowest wavenumbers, and consequently the scale-by-scale balance resembles that of the single-phase case (also included in the figure, see the dashed curves) with minor differences not affecting substantially the resulting scenario.
However, despite the fact that the large-scale dynamics is not affected by the dispersed phase, $F_\mathrm{FIB}$ is active on a broadband high-wavenumber region and responsible of the enhanced small-scale activity, consistently with what observed for the energy spectra.
Overall, at the first wavenumber the energy input $F_\mathrm{FOR}$ is balanced only partially by the dissipation $V$, and the remaining part is transferred by the nonlinear term $T$ to higher wavenumbers. Due to the limited $\Rey$, no constant-flux energy cascade can be observed (i.e., here the flow is multiscale and chaotic but does not exhibit a constant-flux inertial range); nevertheless, a certain proliferation of active scales of motion occurs up to the wavenumber where viscous dissipation becomes dominant.
In contrast, a dramatic change of the overall balance occurs if heavy fibers, or equivalently fixed fibers, are considered (Fig.~\ref{fig:loads_ABC}b). Here, the nonlinear term $T$ turns out to be negligible compared with the other ones. The large-scale dynamics is now governed by the balance between the external forcing $F_\mathrm{FOR}$ and the fluid-structure coupling $F_\mathrm{FIB}$: the former injects energy, while the latter subtracts it. 
All terms are vanishing over an intermediate range of wavenumbers, indicating that the energy transfer occurring in the presence of the dispersed phase is nonlocal. 
Indeed, at large wavenumbers (small scales) the positive contribution of the fluid-structure coupling $F_\mathrm{FIB}$ is essentially balanced by the negative viscous dissipation $V$, with their amplitudes decaying when further increasing $k$.
Notice that the same balance has been observed in Ref.~\cite{gualtieri2017turbulence} for particle-laden homogeneous shear flow. 

These observations show the presence of \added{scale separation due to the fiber addition} and outline the possibility of an effective description for the large-scale dynamics. We therefore propose to model the fluid-structure coupling in Eq.~\eqref{eq:loads} by a Darcy-like friction term $F_\mathrm{FIB}(k) = - \mathcal{D} E(k)$, where $\mathcal{D}$ is the Darcy friction factor. 
The latter can be evaluated by measuring the decay rate of the low-wavenumber components of the spectrum when adding the fibers to the single-phase flow~\footnote{In the Supplementary Materials we report the time history of such decay for a representative case along with the measured values of $\mathcal{D}$.}. The decay of $E(k,t)$ is found to be substantially exponential, so that we can write $E(k,t) = E(k,t=0) \exp(-\beta t) + E(k, t \rightarrow \infty)$, where $\beta = 2 (\mathcal{D} + k^2 \nu)$ is the characteristic time decay rate. Note that this is true as long as the nonlinear term $T$ is negligible. The decay of the first and second spectral modes reveal to be essentially independent of the wavenumber $k$ for all the investigated cases, with $\mathcal D \approx \beta /2$ supporting the validity of the Darcy-like description for the fluid-structure coupling at the large scale.
We observe that the friction factor $\mathcal{D}$ grows with the number of fibers $N$, it is independent of the fibers spatial distribution when isotropic, while it reduces when anisotropic. Finally, $\mathcal{D}$ decreases when fibers are allowed to move.

\begin{figure}
    \centering
    \includegraphics{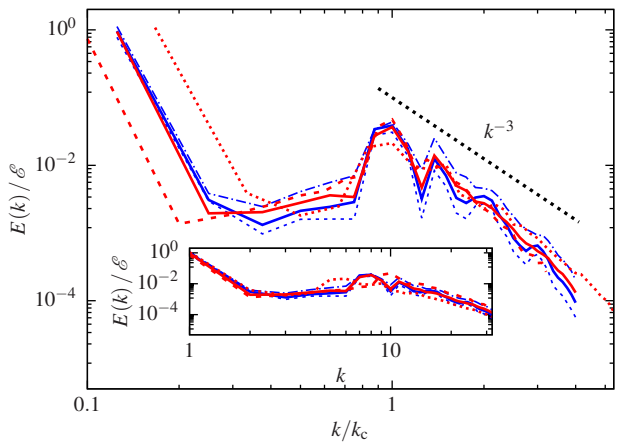}
    \caption{
Normalized energy spectra as a function of the normalized wavenumber where $\mathcal{E} = 3/2\,(\nu/(\nu+\mathcal{D}))^2$ is the kinetic energy of the effective large-scale ABC flow. 
The inset shows the same spectra in the absence of wavenumber rescaling.
Colors and line types are the same as in Fig.~\ref{fig:spectra}.
    }
    \label{fig:spectra_ABC_norm}
\end{figure}

This mean-field large-scale description can be applied to normalize the energy spectra as shown in Fig.~\ref{fig:spectra_ABC_norm} for the different configurations we have analyzed in Figs.~\ref{fig:spectra}a and~\ref{fig:spectra}b. 
Note also that the wavenumber is normalized using the characteristic wavenumber $k_\mathrm{c}$,
so that the small-scale/high-wavenumber peaks for different concentrations substantially overlap (conversely, without normalizing the curves would collapse in the low-wavenumber region, see inset).
The normalization applies over the whole range of scales, confirming the universality of the physical mechanisms under investigation and proving the validity of our phenomenological description.
For the anisotropic case, the Darcy friction features a diagonal tensor $\mathsf{D}$; our analysis applies for each of the three components of $E(k)$ using the corresponding diagonal entry of $\mathsf{D}$, as well as for the total spectrum using the first invariant of $\mathsf{D}$ (i.e., its trace).  
As a final note, the collapsed spectra also reveal a power law $\sim k^{-3}$ scaling at the small scales $k/k_\mathrm{c} \geq 1$, the fingerprint of a regime characterised by smooth fluctuations in space.

\begin{figure}
    \centering
    \includegraphics{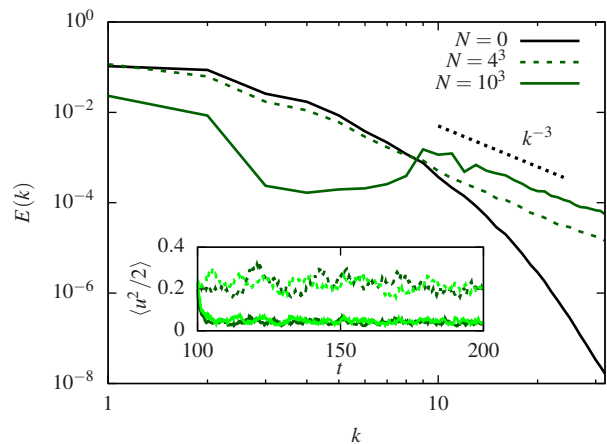}
    \caption{
    Energy spectra for homogeneous isotropic turbulence with fixed fibers of \added{length $c/L=(4\pi)^{-1}$} at different concentrations. The inset shows the time history of the mean fluid kinetic energy for the fully-resolved cases (dark color) and the effective single-phase simulations (light color).
    }
    \label{fig:HIT}
\end{figure}

To further confirm the generality of our findings, we have also considered other flows, such as the parallel Kolmogorov flow (not shown here) and the widely used homogeneous isotropic turbulence (HIT)~\footnote{The flow is sustained by a random forcing in Fourier space on wavenumbers within a shell of radius $k=2$~\cite{eswaran1988forcing,rosti2018flexible}, and the investigated cases are at Taylor's microscale Reynolds number $\Rey_\lambda \approx 40$.}, whose results are reported in Fig.~\ref{fig:HIT}. In particular, the figure shows the energy spectra obtained in HIT for two different fiber concentrations. Again, consistently with what already observed, we find an increase of the small-scale activity (scaling approximately as $k^{-3}$) and a reduction of energy for the large scales. The latter becomes more pronounced as the number of fibers is increased. In the figure, we also present results from effective single-phase computations where we solve the governing equations of the fluid, and the presence of fibers is modelled by the Darcy drag term. Looking at the time history of the mean kinetic energy reported in the inset of Fig.~\ref{fig:HIT}, the effective description provides a very good approximation of the fully-resolved solution, the agreement increasing with the fiber concentration.

A note should be added on the validity of our effective model: the nonlinear contribution to the energy balance has to be negligible.
This is found to occur only when the concentration of the dispersed phase is sufficiently high (although still in the dilute configuration). One additional constraint arises from the fiber inertia in the case of a suspension of moving fibers: the turbulence modulation discussed here is evident only for large enough density ratios, provided the concentration is also high enough.

We have found that a solution of fibers dispersed in turbulence totally changes the classic turbulent energy budget. Due to the back-reaction of the fibers to the flowing fluid, the intensity of the large-scale motion is damped. Flow kinetic energy reduces at large scales being re-injected at small scales causing small-scale mixing. The scenario is robust with respect to changes of the underlying turbulence characteristics. Because we found the same results for both fixed fibers (forming a turbulent porous medium) and freely-moving fibers, the two realms of porous media and suspension dynamics now appear much closer than previously thought. 

\added{To conclude, we expect a similar scenario also for other objects,
such as,  e.g., particles of different shapes or droplets~\cite{dodd_ferrante_2016,lucci_ferrante_elghobashi_2010,gualtieri2017turbulence,rosti_ge_jain_dodd_brandt_2019}.
Our results indeed never appear as consequences of the particular geometry of the immersed objects. Rather, general properties such as their inertia and/or their concentration seem to rule the resulting phenomenology. 
}

\begin{acknowledgments}
SO acknowledges OIST for supporting his visiting period in the Complex Fluids and Flows Unit. 
AM thanks the financial support from the Compagnia di San Paolo, project MINIERA n. I34I20000380007.
LB acknowledges financial support from the Swedish Research Council (VR), Grant No.\ VR 2014-5001. Computing time was provided by INFN-CINECA and SNIC.
\end{acknowledgments}


\providecommand{\noopsort}[1]{}\providecommand{\singleletter}[1]{#1}%

\end{document}